\documentclass[a4paper,10pt]{article}

\usepackage{bm,amsmath,amssymb,graphicx,mathrsfs}

\newcommand{\nc}{\newcommand}
\nc{\rnc}{\renewcommand}
\nc{\nn}{\nonumber}
\nc{\bra}{\langle}
\nc{\ket}{\rangle}
\rnc{\(}{\left(}
\rnc{\)}{\right)}
\rnc{\[}{\left[}
\rnc{\]}{\right]}

\nc{\sh}{\mathrm{sh}}

\begin{document}

\title{A note on a boundary one point function for the six vertex model
with reflecting end}

\author{Kohei Motegi \\
Okayama Institute for Quantum Physics, \\
Kyoyama 1-9-1,
Okayama 700-0015, Japan \\
(e-mail:  motegi@gokutan.c.u-tokyo.ac.jp)}

\maketitle

A boundary one point function related to the 
boundary spontaneous polarization
is studied for the six vertex model
on a $2N \times N$ lattice with
domain wall boundary condition and left reflecting end.
It is expressible in terms of
a special kind of coordinate space
wave functions.
We also express it utilizing determinants.

{\bf Keywords:} boundary six vertex model, one point function,
Yang-Baxter algebra

\section{Introduction}
The six vertex model is 
one of the most fundamental exactly solved models
in statistical physics \cite{Slater,Lieb,Sutherland,Baxter}.
Not only the periodic boundary condition
but also the domain wall boundary condition
is an interesting boundary condition.
For example, the partition function is deeply 
related to the norm \cite{Korepin} and the scalar product
\cite{Slavnov} of the XXZ chain.
The determinant formula of the partition function \cite{Izergin,ICK}
was used to obtain a compact representation of
the scalar product \cite{Slavnov}, which plays a fundamental role
in calculating correlation functions of the XXZ chain
\cite{KIEU,KIB,KMT,KMST}.
The determinant formula was also important for
proving conjectures in enumerative combinatorics 
\cite{Zeilberger,Kuperberg,Bressoud}
such as the numbers of the alternating sign matrices
for a given size.

The calculation of correlation functions are also
interesting in the domain wall boundary condition itself.
Several kinds of them such as the boundary correlation functions
\cite{BKZ,BPZ,FP,CP}
and the emptiness formation probability \cite{CP2} have been calculated.

The mixed boundary conditions of the domain wall and reflecting
boundary \cite{Sklyanin} has also been studied.
The partition function \cite{Tsuchiya}
and several kinds of one point functions \cite{Wang}
are obtained in determinant form.

In this paper, we calculate another kind of
boundary one point function for the six vertex model
on a $2N \times N$ lattice with the mixed boundary condition.
The one point function we consider is different from the ones in
\cite{Wang}.
We show that they can be expressed in terms of
a special kind of coordinate space
wave functions.
Since the coordinate space wave function we use 
can be shown to be expressed as determinants,
we can express the one point function in terms of determinants.

The outline is as follows. We define the six vertex model
with mixed boundary condition in the next section.
In section 3, the one point function is defined and shown to
be expressed in terms of the coordinate space wave function,
and in terms of determinants in section 4.
\section{Six vertex model}
\begin{figure}[htbp]
 \begin{center}
  \includegraphics[width=100mm]{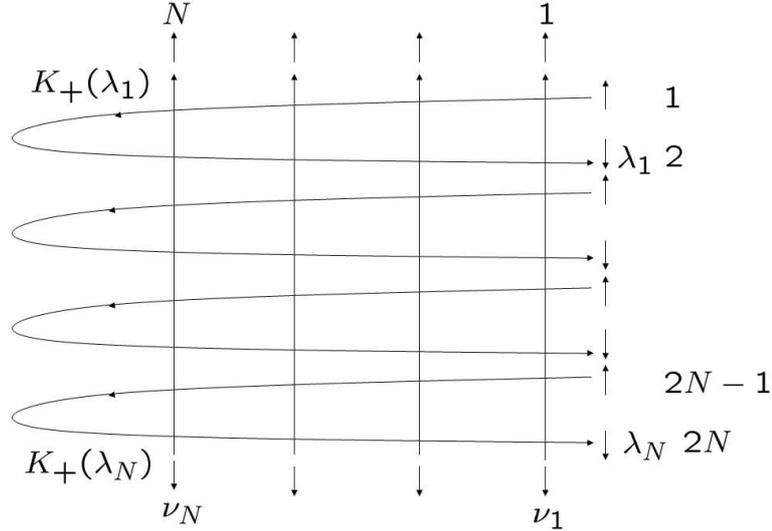}
 \end{center}
 \caption{The six vertex model with left reflecting boundary.}
 \label{fig:one}
\end{figure}

The six vertex model is a model in statistical mechanics,
whose local states
are associated with edges of a square lattice,
which can take two values.
The Boltzmann weights are assigned to its vertices,
and each weight is determined by the configuration
around a vertex. The rule is encoded in the $R$-matrix
\begin{align}
R(\lambda)
=&\left(
\begin{array}{cccc}
a(\lambda)  & 0 & 0 & 0 \\
0 & b(\lambda) & c(\lambda) & 0 \\
0 & c(\lambda) & b(\lambda) & 0 \\
0 & 0 & 0 & a(\lambda)
\end{array}
\right),
\end{align}
where
\begin{align}
a(\lambda)=1, \ b(\lambda)=\frac{ \sh \lambda}{ \sh (\lambda+\eta)},
\ c(\lambda)=\frac{ \sh \eta}{ \sh (\lambda+\eta)}.
\end{align}
The $R$-matrix satisfies the Yang-Baxter equation
\begin{align}
R_{12}(\lambda) R_{13}(\lambda+\mu) R_{23}(\mu)
=R_{23}(\mu) R_{13}(\lambda+\mu) R_{12}(\lambda).
\label{YangBaxter}
\end{align}
In this paper, we consider the six vertex model
on a $2N \times N$ lattice depicted in Figure \ref{fig:one}.
At the upper and lower boundaries, the spins are aligned
all up and all down respectively.
At the right boundary, 
the boundary spins are up for odd rows and down for even rows.
We set $L_{\alpha k}(\lambda_{\alpha},\nu_k)
=R_{\alpha k}(\lambda_{\alpha}-\nu_k-\eta/2)$.
At the intersection of the $\alpha$-th row
and the $k$-th column, we associate the statistical weight
$\sigma_{\beta}^2 L_{\beta k}(-\lambda_{\beta}, \nu_k) \sigma_{\beta}^2,
\ \beta=(\alpha+1)/2$
for $\alpha$ odd and $L_{\beta k}^{t_{\beta}}(\lambda_{\beta},\nu_k),
\ \beta=\alpha/2$
for $\alpha$ even.
Between the $(2 \alpha-1)$-th and $(2 \alpha)$-th row,
the boundary statistical weight
\begin{align}
K_+(\lambda_{\alpha})
=&\left(
\begin{array}{cc}
\sh(\lambda_{\alpha}+\eta/2+\zeta_{+})  & 0 \\
0 & \sh(-\lambda_{\alpha}-\eta/2+\zeta_{+})
\end{array}
\right), \label{Kmatrix}
\end{align}
is associated at the left boundary.
The $K$ matrix \eqref{Kmatrix} satisfies the reflection equation
\begin{align}
&R_{12}(-\lambda_1+\lambda_2) K_+^{1}(\lambda_1)^{t_1}
R_{12}(-\lambda_1-\lambda_2-\eta) K_+^{2}(\lambda_2)^{t_2}
\nonumber \\
&=K_+^{2}(\lambda_2)^{t_2} R_{12}(-\lambda_1-\lambda_2-\eta)
K_+^{1}(\lambda_1)^{t_1} R_{12}(-\lambda_1+\lambda_2).
\end{align}
For convenience, we denote
$\{\lambda  \}=\{ \lambda_1, \lambda_2, \dots, \lambda_N \},
\{ \nu \}=\{ \nu_1, \nu_2, \dots, \nu_N \}$, and
introduce the one-row monodromy matrix
\begin{align}
T(\lambda_{\alpha}, \{ \nu \})=&L_{\alpha N}(\lambda_{\alpha},\nu_N)
\cdots L_{\alpha 1}(\lambda_{\alpha},\nu_1) \nonumber \\
=&\left(
\begin{array}{cc}
A(\lambda_{\alpha}, \{ \nu \})  & B(\lambda_{\alpha}, \{ \nu \}) \\
C(\lambda_{\alpha}, \{ \nu \}) & D(\lambda_{\alpha}, \{ \nu \})
\end{array}
\right). \label{onerow}
\end{align}
Combining two one-row monodromy matrices
and the $K$-matrix \eqref{Kmatrix},
one can construct the double-row monodromy matrix
\begin{align}
U^{t_{\alpha}}(\lambda_{\alpha}, \{ \nu \})&=
T^{t_{\alpha}}(\lambda_{\alpha}, \{ \nu \})K_+(\lambda_{\alpha})
\sigma_{\alpha}^2 T(-\lambda_{\alpha}, \{ \nu \}) \sigma_{\alpha}^2,
\nonumber \\
&=\left(
\begin{array}{cc}
\mathcal{A}(\lambda_{\alpha}, \{ \nu \})  & \mathcal{C}(\lambda_{\alpha}, 
\{ \nu \}) \\
\mathcal{B}(\lambda_{\alpha}, \{ \nu \}) & \mathcal{D}(\lambda_{\alpha},
\{ \nu \})
\end{array}
\right).
\end{align}
The partition function of the six vertex model
with mixed boundary condition, which is the summation
of products of statistical weights over all possible configurations, can
be represented as
\begin{align}
Z_{2N \times N}(\{ \lambda \}, \{ \nu \})
=w_N^- \mathcal{B}(\lambda_{N}, \{ \nu \}) \cdots
\mathcal{B}(\lambda_{1}, \{ \nu \})  w_N^+,
\end{align}
where $w_{N}^+= \prod_{k=1}^{N} \uparrow_k$ and
$w_{N}^-= \prod_{k=1}^{N} \downarrow_k$. It
has the following determinant form \cite{Tsuchiya}
\begin{align}
Z_{2N \times N}(\{ \lambda \}, \{ \nu \})
=\frac{\prod_{j=1}^N \prod_{k=1}^N 
\[ \sh^2(\nu_j+\eta/2)-\sh^2 \lambda_k \]
\det_N \chi(\{ \lambda \}, \{ \nu \})
}
{
\prod_{1 \le j < k \le N}\[ \sh^2 \nu_j -\sh^2 \nu_k \]
\prod_{1 \le j < k \le N}\[ \sh^2 \lambda_k -\sh^2 \lambda_j \]
},
\label{boundarypartition}
\end{align}
where $\chi$ is an $N \times N$ matrix whose elements are given by
\begin{align}
\chi_{jk}&=\chi(\lambda_j, \nu_k), \\
\chi(\lambda, \nu)
&=\frac{-\sh \eta \sh (2 \lambda+\eta) \sh(\nu+\zeta_+)}
{\[ \sh^2(\nu+\eta/2)-\sh^2 \lambda \]
\[  \sh^2(\nu-\eta/2)-\sh^2 \lambda \]}.
\end{align}
\section{One point function}
We consider the following one point function
\begin{align}
F(M)=\frac{f(M)}{Z_{2N \times N}(\{ \lambda \}, \{ \nu \})},
\label{onepoint}
\end{align}
where
\begin{align}
f(M)=w_N^- \mathcal{E}_M(\lambda_N) \mathcal{B}(\lambda_{N-1}, \{\nu \}) \cdots
\mathcal{B}(\lambda_1, \{ \nu \}) w_N^+, \label{numerator}
\end{align}
\begin{align}
\mathcal{E}_M(\lambda_{\alpha})=&
\downarrow_{2 \alpha} V_M(\lambda_{\alpha})
\uparrow_{2 \alpha-1}, \\
V_M(\lambda_{\alpha})=&T^{t_\alpha}(\lambda_\alpha, \{ \nu \})
 \frac{1}{2}(1+\sigma_M^3)
K_+(\lambda_\alpha) 
(\sigma_\alpha^2 L_{\alpha N}(-\lambda_\alpha, \nu_N) \sigma_\alpha^2) \cdots
(\sigma_\alpha^2 L_{\alpha M}(-\lambda_\alpha, \nu_M) \sigma_\alpha^2)
\nonumber \\
&\times
\frac{1}{2}(1+\sigma_\alpha^3)
(\sigma_\alpha^2 L_{\alpha M-1}(-\lambda_\alpha, \nu_{M-1})
\sigma_\alpha^2) \cdots
(\sigma_\alpha^2 L_{\alpha 1}(-\lambda_\alpha, \nu_1) \sigma_\alpha^2),
\end{align}
$w_{N}^+= \prod_{k=1}^{N} \uparrow_k$
and
$w_{N}^-= \prod_{k=1}^{N} \downarrow_k$.
\begin{figure}[htbp]
 \begin{center}
  \includegraphics[width=100mm]{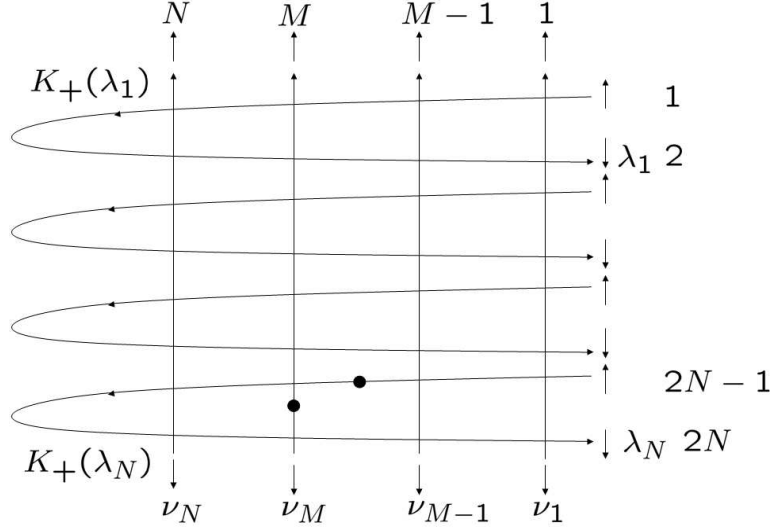}
 \end{center}
 \caption{One point function \eqref{onepoint}.}
 \label{fig:two}
\end{figure}
This one point function is depicted in Figure \ref{fig:two},
and gives the probability that the spin on the $(2N)$-th row 
is turned down just on the $M$-th column. \\
First, with the help of the graphical description of 
the numerator $f(M)$ 
\eqref{numerator}, we find
\begin{align}
f(M)=&\sh(\lambda_N+\eta/2+\zeta_+)
b(-\lambda_N-\nu_M-\eta/2)c(\lambda_N-\nu_M-\eta/2)
\prod_{j=M+1}^N b(\lambda_N-\nu_j-\eta/2) \nonumber \\
&\times \langle 1, \cdots ,\check{M}, \cdots , N|
\prod_{\alpha=1}^{N-1} \mathcal{B}(\lambda_\alpha, \{ \nu \}) w_N^+,
\end{align}
where $\langle x_1, \cdots, x_n |= \ \uparrow_1 \cdots \downarrow_{x_1}
\cdots \downarrow_{x_n} \cdots \uparrow_N$.
Next, utilizing
\begin{align}
\prod_{\alpha=1}^n \mathcal{B}(\lambda_\alpha, \{ \nu \}) w_N^+
=&\prod_{\alpha=1}^n \frac{\sh(2 \lambda_\alpha+\eta)}{\sh (2 \lambda_\alpha)}
\sum_{\sigma_1=\pm} \cdots \sum_{\sigma_n=\pm}
\prod_{\alpha=1}^n \{ (-\sigma_\alpha) \sh(-\sigma_\alpha
\lambda_\alpha+\eta/2-\zeta_+)
d(-\sigma_\alpha \lambda_\alpha, \{ \nu \}) \} \nonumber \\
&\times \prod_{1 \le \alpha < \beta \le n}
\frac{\sh(\sigma_\alpha \lambda_\alpha+\sigma_\beta \lambda_\beta-\eta)}
{\sh(\sigma_\alpha \lambda_\alpha+\sigma_\beta \lambda_\beta)}
\prod_{\alpha=1}^n B(\sigma_\alpha \lambda_\alpha, \{ \nu \}) w_N^+,
\end{align}
where
\begin{align}
d(\lambda, \{ \nu \})=\prod_{j=1}^N b(\lambda-\nu_j-\eta/2),
\end{align}
(see \cite{Wang2} for the rational case), 
one finds \eqref{numerator} can be expressed in
terms of one-row monodromy matrices as
\begin{align}
f(M)=&\sh(\lambda_N+\eta/2+\zeta_+)
b(-\lambda_N-\nu_M-\eta/2)c(\lambda_N-\nu_M-\eta/2)
\prod_{j=M+1}^N b(\lambda_N-\nu_j-\eta/2) \nonumber \\
&\times
\prod_{\alpha=1}^{N-1}
\frac{\sh(2 \lambda_\alpha+\eta)}{\sh (2 \lambda_\alpha)}
\sum_{\sigma_1=\pm} \cdots \sum_{\sigma_{N-1}=\pm}
\prod_{\alpha=1}^{N-1} \{ (-\sigma_\alpha) \sh(-\sigma_\alpha
\lambda_\alpha+\eta/2-\zeta_+)
d(-\sigma_\alpha \lambda_\alpha, \{ \nu \}) \} \nonumber \\
&\times \prod_{1 \le \alpha < \beta \le N-1}
\frac{\sh(\sigma_\alpha \lambda_\alpha+\sigma_\beta \lambda_\beta-\eta)}
{\sh(\sigma_\alpha \lambda_\alpha+\sigma_\beta \lambda_\beta)}
\langle 1, \cdots ,\check{M}, \cdots , N|
\prod_{\alpha=1}^{N-1} B(\sigma_\alpha \lambda_\alpha, \{ \nu \}) w_N^+.
\label{expressionpre}
\end{align}
$\langle 1, \cdots ,\check{M}, \cdots , N|
\prod_{\alpha=1}^{N-1} B(\sigma_\alpha \lambda_\alpha, \{ \nu \}) w_N^+$,
which appears in the last equation, is a special case of the
coordinate space wave function \cite{Yang,Gaudin,Ovchinnikov,AL,GM,KM}
\begin{align}
\psi(x_1, \cdots , x_n)=
\langle x_1, \cdots , x_n|
\prod_{\alpha=1}^{n} B(\lambda_\alpha, \{ \nu \}) w_N^+,
\end{align}
whose expression is given by
\begin{align}
\psi(x_1, \cdots , x_n)
=\sum_{P \in S_n}A(P, \{ \lambda_1, \dots, \lambda_n \})
\prod_{j=1}^n \phi_{P_j}(x_j, \{ \lambda_1, \dots, \lambda_n \}),
\label{coordinate}
\end{align}
where $S_n$ is the symmetric group of order $n$ and
\begin{align}
A(P, \{ \lambda_1, \dots, \lambda_n \})
&=\prod_{1 \le \alpha < \beta \le n}
b^{-1}(\lambda_{P_\beta}-\lambda_{P_\alpha}), \\
\phi_{P_j}(x_j, \{ \lambda_1, \dots, \lambda_n \})&=
c(\lambda_{P_j}-\eta/2-\nu_{x_j})
\prod_{1 \le l < x_j}b(\lambda_{P_j}-\nu_l-\eta/2),
\end{align}
where $P$ is an element of $S_n$.
Thus, from \eqref{expressionpre} and \eqref{coordinate},
one has the explicit expression of \eqref{onepoint}
\begin{align}
F(M)=&
\frac{\sh \eta \
\sh(\lambda_N+\eta/2+\zeta_+)
\sh(\lambda_N+\nu_M+\eta/2)}
{Z_{2N \times N}(\{ \lambda \}, \{ \nu \})
[ \sh^2 \lambda_N
-\sh^2 (\nu_M-\eta/2)]}
\prod_{j=M+1}^N 
\frac{\sh(\lambda_N-\nu_j-\eta/2)}
{\sh(\lambda_N-\nu_j+\eta/2)}
\nonumber \\
&\times
\prod_{\alpha=1}^{N-1}
\frac{\sh(2 \lambda_\alpha+\eta)}{\sh (2 \lambda_\alpha)}
\sum_{\sigma_1=\pm} \cdots \sum_{\sigma_{N-1}=\pm} \sum_{P \in S_{N-1}}
\prod_{\alpha=1}^{N-1} \{ (-\sigma_\alpha) \sh(-\sigma_\alpha
\lambda_\alpha+\eta/2-\zeta_+) \} \nonumber \\
&\times \prod_{\alpha=1}^{N-1} \prod_{k=1}^N
\frac{\sh(-\sigma_\alpha \lambda_\alpha-\nu_k-\eta/2)}
{\sh(-\sigma_\alpha \lambda_\alpha-\nu_k+\eta/2)}
\prod_{1 \le \alpha < \beta \le N-1}
\frac{\sh(\sigma_\alpha \lambda_\alpha+\sigma_\beta \lambda_\beta-\eta)}
{\sh(\sigma_\alpha \lambda_\alpha+\sigma_\beta \lambda_\beta)}
\nonumber \\
&\times
A(P, \sigma(\{ \lambda \} \backslash \lambda_N))
\prod_{j=1}^{M-1}\phi_{P_j}(j, \sigma(\{ \lambda \} \backslash \lambda_N)  )
\prod_{j=M}^{N-1}\phi_{P_j}(j+1,  \sigma(\{ \lambda \} \backslash \lambda_N) ),
\end{align}
where $\sigma(\{ \lambda \} \backslash \lambda_N)=\{
\sigma_1 \lambda_1, \sigma_2 \lambda_2, \dots, \sigma_{N-1} \lambda_{N-1} \}$.
\section{Determinant representation}
One can also directly express the
coordinate space wave functions in \eqref{expressionpre}
as deteminants. We change the
viewpoint to use the column monodromy matrix
\begin{align}
\overline{T}(\nu_j, \{ \lambda \} \backslash \lambda_N)
=&L_{N-1j}(\lambda_{N-1}, \nu_j)
\cdots L_{1j}(\lambda_{1}, \nu_j) \nonumber \\
=&\left(
\begin{array}{cc}
\overline{A}(\nu_j, \{ \lambda \} \backslash \lambda_N) &
\overline{B}(\nu_j, \{ \lambda \} \backslash \lambda_N) \\
\overline{C}(\nu_j, \{ \lambda \} \backslash \lambda_N) &
\overline{D}(\nu_j, \{ \lambda \} \backslash \lambda_N)
\end{array}
\right),
\end{align}
instead of the row transfer matrix \eqref{onerow}.
The coordinate space wave function which we should consider
can be expressed in terms of column monodromy matrix as 
\begin{align}
&\langle 1, \cdots ,\check{M}, \cdots , N|
\prod_{\alpha=1}^{N-1} B(\lambda_\alpha, \{ \nu \}) w_N^+
\nonumber \\
=&v_{N-1}^+ \overline{C}(\nu_N, \{ \lambda \} \backslash \lambda_N)
\cdots \overline{C}(\nu_{M+1}, \{ \lambda \} \backslash \lambda_N)
\overline{A}(\nu_M, \{ \lambda \} \backslash \lambda_N)
\overline{C}(\nu_{M-1}, \{ \lambda \} \backslash \lambda_N)
\cdots \overline{C}(\nu_{1}, \{ \lambda \} \backslash \lambda_N) v_{N-1}^-
\nonumber \\
=&v_{N-1}^- \overline{B}(\nu_N, \{ \lambda \} \backslash \lambda_N)
\cdots \overline{B}(\nu_{M+1}, \{ \lambda \} \backslash \lambda_N)
\overline{D}(\nu_M, \{ \lambda \} \backslash \lambda_N)
\overline{B}(\nu_{M-1}, \{ \lambda \} \backslash \lambda_N)
\cdots \overline{B}(\nu_{1}, \{ \lambda \} \backslash \lambda_N) v_{N-1}^+,
\label{change}
\end{align}
where
$v_{N-1}^+= \prod_{\alpha=1}^{N-1} \uparrow_\alpha$
and
$v_{N-1}^-= \prod_{\alpha=1}^{N-1} \downarrow_\alpha$.
One can express \eqref{change} in determinant form
(cf. \cite{BPZ}) utilizing
\begin{align}
&c(\nu-\mu)\overline{B}(\mu, \{ \lambda \} \backslash \lambda_N)
\overline{D}(\nu, \{ \lambda \} \backslash \lambda_N)
+b(\nu-\mu)\overline{D}(\mu, \{ \lambda \} \backslash \lambda_N)
\overline{B}(\nu, \{ \lambda \} \backslash \lambda_N) \nonumber \\
&=\overline{B}(\nu, \{ \lambda \} \backslash \lambda_N)
\overline{D}(\mu, \{ \lambda \} \backslash \lambda_N),
\label{onerelation}
\end{align}
which follows from the Yang-Baxter equation \eqref{YangBaxter}
\begin{align}
R_{12}(\nu-\mu)\overline{T_1}(\mu, \{ \lambda \} \backslash \lambda_N)
\overline{T_2}(\nu, \{ \lambda \} \backslash \lambda_N)
=\overline{T_2}(\nu, \{ \lambda \} \backslash \lambda_N)
\overline{T_1}(\mu, \{ \lambda \} \backslash \lambda_N)
R_{12}(\nu-\mu),
\end{align}
the action of $\overline{D}$ on the vacuum
\begin{align}
\overline{D}(\nu, \{ \lambda \} \backslash \lambda_N) v_{N-1}^+
=\prod_{\alpha=1}^{N-1}b(\lambda_\alpha-\nu-\eta/2) v_{N-1}^+,
\end{align}
and the determinant representation of the 
partition function of the six vertex model on a $n \times n$
lattice with domain wall boundary condition
\begin{align}
v_n^- \prod_{j=1}^n \overline{B}(\nu_j, \{ \lambda_1 , \dots, \lambda_n \}) v_n^+
=\frac{\prod_{\alpha,k=1}^n \sh(\lambda_\alpha-\nu_k-\eta/2)
\det_n M
}
{\prod_{1 \le j < k \le n} \sh(\nu_j-\nu_k)
\prod_{1 \le \alpha < \beta \le n} \sh(\lambda_\beta-\lambda_\alpha)},
\end{align}
where $M$ is an $n \times n$ matrix whose elements are
\begin{align}
M_{\alpha k}&=\phi(\lambda_\alpha, \nu_k), \\
\phi(\lambda, \nu)
&=\frac{\sh \eta}{\sh(\lambda-\nu+\eta/2)
\sh(\lambda-\nu-\eta/2)}.
\end{align}
The result is
\begin{align}
&v_{N-1}^- \overline{B}(\nu_N, \{ \lambda \} \backslash \lambda_N)
\cdots \overline{B}(\nu_{M+1}, \{ \lambda \} \backslash \lambda_N)
\overline{D}(\nu_M, \{ \lambda \} \backslash \lambda_N)
\overline{B}(\nu_{M-1}, \{ \lambda \} \backslash \lambda_N)
\cdots \overline{B}(\nu_{1}, \{ \lambda \} \backslash \lambda_N ) v_{N-1}^+
\nonumber \\
&=\frac{\prod_{\alpha=1}^{N-1} \prod_{k=1}^N
\sh(\lambda_\alpha-\nu_k-\eta/2)
\det_N H(\{ \nu \}, \{ \lambda \} \backslash \lambda_N)}
{\prod_{1 \le j < k \le N} \sh(\nu_k-\nu_j)
\prod_{1 \le \alpha < \beta \le N-1} \sh(\lambda_\alpha-\lambda_\beta)}.
\label{det}
\end{align}
Here, $H$ is an $N \times N$ matrix whose elements are given by
\begin{align}
H_{1k}&=h(\nu_k, \{ \lambda \} \backslash \lambda_N), \\
H_{\alpha k}&=\phi(\lambda_{\alpha-1}, \nu_k), \ \alpha \ge 2,
\end{align}
where
\begin{align}
h(\nu, \{ \lambda \} \backslash \lambda_N)&=
\frac{\prod_{k=1}^{M-1} \sh(\nu_k-\nu+\eta)
\prod_{k=M+1}^N \sh(\nu_k-\nu)}
{\prod_{\alpha=1}^{N-1} \sh(\lambda_\alpha-\nu+\eta/2)}.
\end{align}
Combining \eqref{expressionpre}, \eqref{change} and
\eqref{det}, we have
\begin{align}
f(M)=&
\frac{\sh \eta \
\sh(\lambda_N+\eta/2+\zeta_+)
\sh(\lambda_N+\nu_M+\eta/2)}
{[ \sh^2 \lambda_N -
\sh^2 (\nu_M-\eta/2) ]}
\prod_{j=M+1}^N 
\frac{\sh(\lambda_N-\nu_j-\eta/2)}
{\sh(\lambda_N-\nu_j+\eta/2)}
\nonumber \\
&\times
\prod_{\alpha=1}^{N-1}
\frac{\sh(2 \lambda_\alpha+\eta)}{\sh (2 \lambda_\alpha)}
\sum_{\sigma_1=\pm} \cdots \sum_{\sigma_{N-1}=\pm}
\prod_{\alpha=1}^{N-1} \{ (-\sigma_\alpha) \sh(-\sigma_\alpha
\lambda_\alpha+\eta/2-\zeta_+) \} \nonumber \\
&\times \prod_{\alpha=1}^{N-1} \prod_{k=1}^N
\frac{\sh(-\sigma_\alpha \lambda_\alpha-\nu_k-\eta/2)}
{\sh(-\sigma_\alpha \lambda_\alpha-\nu_k+\eta/2)}
\prod_{1 \le \alpha < \beta \le N-1}
\frac{\sh(\sigma_\alpha \lambda_\alpha+\sigma_\beta \lambda_\beta-\eta)}
{\sh(\sigma_\alpha \lambda_\alpha+\sigma_\beta \lambda_\beta)}
\nonumber \\
&\times
\frac{\prod_{\alpha=1}^{N-1} \prod_{k=1}^N
\sh(\sigma_\alpha \lambda_\alpha-\nu_k-\eta/2)
\det_N H(\{ \nu \}, \sigma(\{ \lambda \} \backslash \lambda_N))}
{\prod_{1 \le j < k \le N} \sh(\nu_k-\nu_j)
\prod_{1 \le \alpha < \beta \le N-1} \sh(\sigma_\alpha \lambda_\alpha
-\sigma_\beta \lambda_\beta)}. \label{numeratorexpression}
\end{align}
Dividing \eqref{numeratorexpression} by the partition function
\eqref{boundarypartition} and simplifying,
we can express \eqref{onepoint}
as a sum of determinants as
\begin{align}
F(M)=&
\frac{\sh \eta \
\sh(\lambda_N+\eta/2+\zeta_+)
\sh(\lambda_N+\nu_M+\eta/2)}
{[ \sh^2 \lambda_N -
\sh^2 (\nu_M-\eta/2) ]}
\prod_{j=1}^{N-1}\frac{\sh(2\lambda_j+\eta)}{\sh 2 \lambda_j} \nonumber \\
&\times 
\frac{\prod_{1 \le j < k \le N}\sh(\nu_j+\nu_k)
\prod_{j=1}^{N-1}[\sh^2 \lambda_j-\sh^2 \lambda_N ]}
{\prod_{j=1}^M [ \sh^2 (\nu_j+\eta/2)-\sh^2 \lambda_N ]
\prod_{j=M+1}^N [ \sh^2 \nu_j -\sh^2 (\lambda_N+\eta/2)]
} \nonumber \\
&\times \sum_{\sigma_1=\pm} \cdots \sum_{\sigma_{N-1}=\pm}
\prod_{\alpha=1}^{N-1} \{ (-\sigma_\alpha) \sh(-\sigma_\alpha
\lambda_\alpha+\eta/2-\zeta_+) \} \nonumber \\
&\times
\frac{\prod_{1 \le \alpha < \beta \le N-1}
\sh(\sigma_\alpha \lambda_\alpha+\sigma_\beta \lambda_\beta-\eta)
}
{
\prod_{\alpha=1}^{N-1} \prod_{k=1}^N
\sh(-\sigma_\alpha \lambda_\alpha-\nu_k+\eta/2)
}
\frac{
\det_N H(\{ \nu \}, \sigma(\{ \lambda \} \backslash \lambda_N))
}
{
\det_N \chi(\{ \lambda \}, \{ \nu \})
}.
\end{align}

\section{Conclusion}
In this paper, we considered a kind of one point function
for the six vertex model with domain wall boundary condition and
left reflecting boundary. We showed  that it can be expressed in terms
of a certain kind of coordinate space wave function. Since the coordinate 
space wave functions we use can be shown to be expressed as determinants,
the one point function can also be expressed as combinations of
determinants.

It is interesting to extend the analysis to two point functions.
For example, one can consider the probability of finding certain 
configurations around two vertices at the lower and right boundaries. 
It is intriguing to calculate these kinds of correlation functions.


\section*{References}

\begin{thebibliography}{00}
%
\bibitem{Slater}
J.C.~Slater: {\it J. Chem. Phys.} {\bf 9}, 16 (1941).
%
\bibitem{Lieb}
E.H.~Lieb: {\it Phys. Rev.} {\bf 162}, 162 (1967).
%
\bibitem{Sutherland}
B.~Sutherland: {\it Phys. Rev. Lett.} {\bf 19}, 103 (1967).
%
\bibitem{Baxter}
R.J.~Baxter: {\it Exactly Solved Models in Statistical Mechanics},
Academic press, San Diego 1982.
%
\bibitem{Korepin}
V.E.~Korepin: {\it Commun. Math. Phys.} {\bf 86}, 391 (1982).
\bibitem{Slavnov}
N.A.~Slavnov: {\it Theor. Math. Phys.} {\bf 79}, 502 (1989).
%
\bibitem{Izergin}
A.G.~Izergin: {\it Sov. Phys. Dokl.} {\bf 32}, 878 (1987).
%
\bibitem{ICK}
A.G.~Izergin, D.A.~Coker and V.E.~Korepin:
{\it J. Phys. A} {\bf 25}, 4315 (1992). 
%
\bibitem{KIEU}
V.E.~Korepin, A.G.~Izergin, F.H.L~Essler
and D.B.~Uglov:
{\it Phys. Lett. A} {\bf 190}, 182 (1994).
%
\bibitem{KIB}
V.E.~Korepin, N.M.~Bogoliubov and A.G.~Izergin:
{\it Quantum Inverse Scattering Method and Correlation Functions},
Cambridge University Press, Cambridge 1993.
%
\bibitem{KMT}
N.~Kitanine, J.M.~Maillet and V.~Terras:
{\it Nucl. Phys. B} {\bf 567}, 554 (2000). 
\bibitem{KMST}
N.~Kitanine, J.M.~Maillet, N.A.~Slavnov and V.~Terras:
{\it Nucl. Phys. B} {\bf 641}, 487 (2002).
%
\bibitem{Zeilberger}
D.~Zeilberger:
{\it Elec. J. Comb.} {\bf 3}, (2) R13 (1996). 
%
\bibitem{Kuperberg}
G.~Kuperberg: {\it Int. Math. Res. Not.} {\bf 1996},
139 (1996).
%
\bibitem{Bressoud}
D.M.~Bressoud:
{\it Proofs and Confirmations: The Story of the 
Alternating Sign Matrix Conjecture},
Cambridge University Press, Cambridge 1999.
%
\bibitem{BKZ}
N.M.~Boboliubov, A.V.~Kitaev and M.B.~Zvonarev:
{\it Phys. Rev. E} {\bf 65}, 026126 (2002).
%
\bibitem{BPZ}
N.M.~Bogoliubov, A.G.~Pronko and M.B.~Zvonarev:
{\it J. Phys. A} {\bf 35}, 5525 (2002).
%
\bibitem{FP}
O.~Foda and I.~Preston:
{\it J. Stat. Mech.} P11001 (2004).
%
\bibitem{CP}
F.~Colomo and A.G.~Pronko:
{\it J. Stat. Mech.} P05010 (2005).
%
\bibitem{CP2}
F.~Colomo and A.G.~Pronko:
{\it Nucl. Phys. B} {\bf 798}, 340 (2008).
%
\bibitem{Sklyanin}
E.K.~Sklyanin:
{\it J. Phys. A} {\bf 21}, 2375 (1998).
%
\bibitem{Tsuchiya}
O.~Tsuchiya:
{\it J. Math. Phys.} {\bf 39}, 5946 (1998).
%
\bibitem{Wang}
Y-S.~Wang:
{\it J. Phys. A} {\bf 36}, 4007 (2003).
%
\bibitem{Wang2}
Y-S.~Wang: 
{\it Nucl. Phys. B} {\bf 622}, 633 (2002).
%
\bibitem{Yang}
C.N.~Yang:
{\it Phys. Rev. Lett.} {\bf 19}, 1312 (1967).
%
\bibitem{Gaudin}
M.~Gaudin:
{\it Phys. Lett. A} {\bf 24}, 55 (1967).
%
\bibitem{Ovchinnikov}
A.A.~Ovchinnikov:
{\it Phys. Lett. A} {\bf 374}, 1311 (2010).
%
\bibitem{AL}
F.C.~Alcaraz and M.J.~Lazo:
{\it J. Phys. A} {\bf 39}, 11335 (2006).
\bibitem{GM}
O.~Golinelli and K.~Mallick:
{\it J. Phys. A} {\bf 39}, 10647 (2006).
\bibitem{KM}
H.~Katsura and I.~Maruyama:
{\it J. Phys. A} {\bf 43}, 175003 (2010).
%

\end{thebibliography}
\end{document}